\documentclass{PoS}

\newcommand{\alxxvi}{$^{26}$Al\ }

\newcommand{\sixxviii}{$^{28}$Si\ }
\newcommand{\sxxxii}{$^{32}$S\ }
\newcommand{\oxvi}{$^{16}$O\ }
\newcommand{\fxviii}{$^{18}$F\ }
\newcommand{\nxiv}{$^{14}$N\ }
  
\newcommand{\sol}{$M_\odot$\,}
\def \nuc#1#2{\relax\ifmmode{}^{#1}{\protect\text{#2}}\else${}^{#1}$#2\fi}

\title{Spatial Distribution of Nucleosynthesis Products in Cassiopeia A: Comparison Between Observations and 3D Explosion Models}

\ShortTitle{Spatial Distribution of Nucleosynthesis Products in Cassiopeia A: Comparison Between Observations and 3D Explosion Models}

\author{\speaker{Patrick Young}\\
       Arizona State University \\
       E-mail: \email{patrick.young.1@asu.edu}}

\author{{Carola I. Ellinger, Frank Timmes}\\
        Arizona State University\\}
        
\author{{David Arnett}\\
       University of Arizona\\}
              
\author{{Chris L. Fryer, Gabriel Rockefeller, Aimee Hungerford, \& Steven Diehl}\\
        Los Alamos National Laboratory\\}
        
\author{{Michael Bennett, Raphael Hirschi, \& Marco Pignatari}\\
        Keele University\\}
        
\author{{Falk Herwig}\\
        University of Victoria\\ }       
        
\author{{Georgios Magkotsios}\\
	University of Notre Dame\\  }     

\abstract{We examine observed heavy element abundances in the Cassiopeia A
supernova remnant as a constraint on the nature of the Cas A
supernova. We compare bulk abundances from 1D and 3D explosion models
and spatial distribution of elements in 3D models with those derived
from X-ray observations. We also examine the cospatial production of \alxxvi with other species. We find that the most reliable indicator of the presence of \alxxvi in unmixed ejecta is a very low S/Si ratio ($\sim 0.05$). Production of N in O/S/Si-rich regions is also indicative. The biologically important element P is produced at its highest abundance in the same regions. Proxies should be detectable in supernova ejecta with high spatial resolution multiwavelength observations.}

\FullConference{10th Symposium on Nuclei in the Cosmos\\
		 July 27 - August 1 2008\\
		 Mackinac Island, Michigan, USA}

\begin{document}

\section{Introduction}

Cassiopeia A is perhaps the best studied young Galactic supernova
Remnant (SNR). It is nearby (3.4 kpc) \cite{reed95} and young ($\sim
325$ yr) \cite{thor01}. The wealth of data from ground-based
observations in the optical, IR, and radio and from space in the
optical, x-ray, and $\gamma$-ray allow us to study its morphology and
composition in great detail. 
In \cite{young06a} we attempted to narrow down the possibilities for
the progenitor star of Cas A using comparisons between computational
models and different lines of observational evidence. We believe the
most probable candidate for the progenitor of Cas A is a 15-25 \sol
star in an unequal mass binary. This will form the basic assumption of
our analysis in this paper.

This work ties into the larger NuGrid collaboration. We will examine the sensitivity of our results to our choice of nuclear rates , determining the dominant rates affecting these yields and calculating the error in these yields caused by the uncertainty in the rates. We can then compare these errors to current errors in the hydrodynamics, using observed supernovae such as Cas A as verification tests in the manner discussed here. 

\section{Calculations}
These calculations explore four different progenitor models each for a range of explosion scenarios. We use a large set of thermally driven 1D explosions with varying delays for a star of initial mass 23 \sol and a more restricted range of explosions for a 16 \sol and 23 \sol with the hydrogen envelope stripped in a case B binary scenario and a 40 \sol that ends its life as a type WC/O after extensive mass loss. We also examine a 3D explosion of the 23 \sol binary progenitor. 

\subsection{1D Explosions}

To model collapse and explosion, we use a 1-dimensional Lagrangian
code developed by Herant et al. (1994).  This code includes 3-flavor
neutrino transport using a flux-limited diffusion calculation and a
coupled set of equations of state to model the wide range of densities
in the collapse phase \cite{her94,fryer99a}. It
includes a 14-element nuclear network \cite{bth89} to follow
energy generation.  

\subsection{3D Explosions \label{3dcalc}}

Our 3-dimensional simulations use the output of the 1-D explosion (23\,M$_\odot$ star, 23m-run5) when the shock has reached 10$^9$\,cm.  We map the structure of this explosion
into our 3D Smooth Particle Hydrodynamics code SNSPH \cite{frw06}.  To simulate an asymmetric explosion,
we modify the velocities within each shell based on angular position.  The velocities of particles within 30$^\circ$ of the
z-axis were increased by a factor of 6 and the remaining particles were decreased by a factor of 1.2, roughly conserving
the explosion energy.  We will refer to these as high velocity structures (HVS).  At these early times in the
explosion, much of the energy remains in thermal energy, so the total asymmetry in the explosion is not
as extreme as our velocity modifications suggest.  The large velocity asymmetry results in roughly a factor of two spatial asymmetry between the axes.  In this calculation, we model the explosion using 1 million SPH particles.

\subsection{Nucleosynthesis Post-processing}

Nucleosynthesis
post-processing was performed with the Burn code \cite{yf07}, using
a 524 element network terminating at $^{99}$Tc.  Reverse rates are
calculated from detailed balance and allow a smooth transition to a
nuclear statistical equilibrium (NSE) solver at $T > 10^{10}$K. 

\section{Explosion Geometry}

At 800 seconds we cannot yet compare the simulation to the detailed observed abundances, but we can make statements about the gross morphology. We must account for two features. Most of the material is in a circular, flattened structure that is close to the plane of the sky. This is well-delineated by the fast moving optical knots, which are largely absent from the center of the remnant. In the x-ray, we see that higher atomic number species exist preferentially towards the center of the remnant \cite{wil02}. These features can be reproduced at least qualitatively with a bipolar explosion with the axis oriented close to our line of sight. When the HVS expand the bubbles "rupture" after leaving the dense remnant, producing a flattened main remnant with nucleosynthesis products of higher A progressively closer to the symmetry axis.

\section{\alxxvi Production \label{Geometry}}

\cite{odh07} find that for uniformly expanding SN ejecta, the reverse shock caused by the initial interaction of ejecta and a protoplanetary disk develops into a bow shock that deflects most of the supernova material around the disk, with injection efficiencies of order of a percent. The average ejecta composition is therefore probably not the appropriate quantity to add to the disk when considering enrichment by \alxxvi. An ejecta knot with the typical characteristics of those in the Cassiopeia A SNR would have a mostly or completely unmixed composition, and could deliver pure \alxxvi -rich material to a disk. Dense optical and infrared knots have densities at least three orders of magnitude greater than uniformly expanding ejecta.  A dense knot is also the most likely part of the ejecta to undergo dust condensation. \cite{odh08}  find that for dust particles with a diameter $> 0.01 {\rm \mu m}$ the injection efficiencies approach 100\%. For generous assumptions of the mass of ejected metals, distance from a supernova, and covering factor of a protoplanetary disk we estimate that a disk is likely to encounter only  of order one knot that contributes significantly to its mass. Since \alxxvi or its proxies are unlikely to be detectable in a stellar/planetary system, we will consider looking for suitable enriching material in supernova remnants.

\subsection{Detectability of \alxxvi\ Proxies}

Can actually detect the characteristic enhancement or ratios of nucleosynthetic products that can reliably indicate the presence of \alxxvi in supernova ejecta?
\begin{figure}
\includegraphics[scale=0.3]{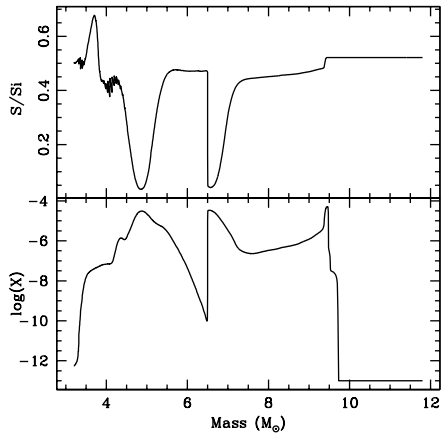}
\caption{S/Si (top) and \alxxvi\ (bottom) for model 23m-run5. The ratio drops in the regions of highest \alxxvi production. Other explosions show a similar pattern. }
\label{ssi.fig}
\end{figure}
Si and S abundances turn out to be highly diagnostic (Figure~\ref{ssi.fig}) because of the production channel. The burning raises the Si abundance from log(X) $\sim -3$ to $\sim -1.5$, primarily in \sixxviii. This is the typical abundance in the high \alxxvi\  regions. Temperatures are not high enough to produce significant amounts of S, so S/Si drops by a factor of about  10. Oxygen burning produces S efficiently at low entropies (\oxvi (\oxvi ,$\gamma$)\sxxxii ) and at high T transitioning to QSE, where S is thermodynamically favored over Si. As a result there is only a narrow range of mass in the star where S/Si drops much below 0.5. This coincides with the peak mass fraction of \alxxvi.

\begin{figure}
\includegraphics[scale=0.3]{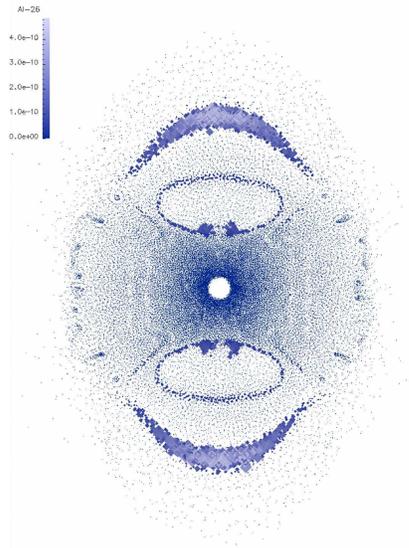}
\caption{\alxxvi\  mass per particle for a $10^9$cm thick slice through the x-z plane. The outer bubble of high \alxxvi\ abundance has reached C burning temperatures. The inner ring and bubble reach Ne burning conditions. }
\label{al26.fig}
\end{figure}

We find a similar pattern in the 3D explosion to the 1D results (Figure~\ref{al26.fig}. In this case, in the spherically symmetric region no material is processed at high enough temperatures to produce \alxxvi. The HVS have a large region that consists of material that underwent carbon burning in the progenitor star at temperatures of $\sim 8\times 10^8$K and reached peak shock temperatures of between 1 and 2$\times 10^9$K. This is the \alxxvi\ bubble. The ring consists of material that undergoes Ne burning in the progenitor ($1-1.5\times 10^9$K) and explosive Ne and/or C  burning ($2.2-2.8\times 10^9$K).  As in the 1D explosions, S/Si is very low and P production is high. The 3D explosion, with its more complicated thermodynamic history, reminds us that 1D does not tell the whole story. The yield is increased by higher temperatures and higher densities, but \alxxvi\ is rapidly destroyed by succeeding reactions. if there is a freezeout after the shock in which density drops rapidly, the high production can be achieved without subsequent destruction. This is the reason for the higher \alxxvi\ production in the sub-explosive C burning bubble than the explosive ring, the opposite of the pattern found in 1D. Large amounts of \fxviii are produced under the same conditions. If this excess flourine is not destroyed by subsequent burning it preferentially decays by \fxviii($\gamma, \alpha$)\nxiv at temperatures above T$\sim 1\times 10^9$K. Thus we have materialwith high levels of both N and O, Si, and S. These are potentially analogous to the Mixed Emission Knots (MEKs) identified by \cite{fesen01} in Cas A.

Detection of \alxxvi-rich material in core-collapse supernova remnants is a favorable situation. The characteristically low S/Si will be undiluted by mixing with ISM material in young remnants, so we should be able to predict which ejecta knots have high \alxxvi\ in well-resolved cases like Cas A. O-rich knots with high Mg or Na or moderately enhanced Si are good candidates for a combined optical/IR survey. The S/Si ratio of these knots would then provide a strong test of whether they would contain \alxxvi at high abundance. The Mixed Emission Knots in Cas A may also be excellent candidates. These knots are unique due to the simultaneous presence of significant emission from N and from O and S. The burning stages in a star that produce O and S destroy N very efficiently. It has therefore been assumed that these knots are most likely material from different locations in the progenitor star that are superimposed on line of sight or somehow mixed during the explosion.  If these clumps are produced by a nucleosynthetic process they are most likely produced in the same conditions favorable for  \alxxvi production.

\end{document}